\documentclass[reprint, superscriptaddress, amsmath,amssymb, aps, prresearch, longbibliography, floatfix]{revtex4-2}
\usepackage{amsmath}
\usepackage{graphicx}
\usepackage[normalem]{ulem}
\usepackage{bm}
\usepackage{natbib}
\usepackage{dsfont}
\usepackage{tikz}
\usepackage{epsfig}
\usepackage{feynmf}
\usepackage{blindtext, rotating}
\usepackage{mathtools}
\usepackage{dsfont}
\usepackage{subcaption}
\usepackage{physics}
\usepackage{amsfonts}
\usepackage{xcolor}
\usepackage{ragged2e}
\usepackage{siunitx}
\usepackage{comment}
\usepackage{soul}
\usepackage{empheq}
\usepackage{lipsum}
\usepackage{hyperref} 
\hypersetup{breaklinks=true, colorlinks=true, citecolor=blue, linkcolor=cyan, urlcolor=blue,filecolor=blue}

\usepackage[T1]{fontenc}

\usepackage{xcolor} 

\DeclareCaptionJustification{justified}{\justifying}

\DeclareSIUnit{\rad}{rad}

\definecolor{bright_blue}{HTML}{85C1E9}
\definecolor{middle_blue}{HTML}{2E86C1}
\definecolor{dark_blue}{HTML}{1B4F72}

\begin{document}

\title{Entanglement and squeezing of gravitational waves}

\author{Thiago Guerreiro}
\email{barbosa@puc-rio.br}
\affiliation{Department of Physics, Pontifical Catholic University of Rio de Janeiro, Rio de Janeiro 22451-900, Brazil}

\begin{abstract}
We show that the self-interactions present in the effective field theory formulation of general relativity can couple gravitational wave modes and generate nonclassical states. The output of gravitational nonlinear processes can also be sensitive to quantum features of the input states, indicating that nonlinearities can act both as sources and detectors of quantum features of gravitational waves. Due to gauge and quantization issues in strongly curved spacetimes, we work in the geometric optics limit of gravitational radiation, but we expect the key ideas extend to situations of astrophysical interest.
This offers a new direction for probing the quantum nature of gravity, analogous to how the quantumness of electrodynamics was established through quantum optics. 

\end{abstract}


\maketitle


\textit{Introduction.}---  Currently, no observed phenomenon requires a quantum description of gravity \cite{unruh1984steps}. Yet, at low energies the effective field theory quantization of general relativity forecasts a number of features departing from the classical theory \cite{donoghue1994general}. Recently, various proposals aimed at testing some of these predictions have been put forward, mainly following \cite{marletto2017gravitationally, bose2017spin, krisnanda2020observable, qvarfort2020mesoscopic}. 
On one hand, we have tabletop experiments seeking to prepare macroscopic quantum superpositions of massive objects \cite{aspelmeyer2022avoid, bose2025massive, marletto2025quantum}. From large superposition states we could witness the quantum nature of gravity through gravitationally induced entanglement due to the exchange of virtual quanta \cite{carney2022newton, belenchia2018quantum, danielson2022gravitationally}. Alternatively, we may look for tree-level quantum effects in gravitational waves (GWs) \cite{guerreiro2020gravity, chawla2023quantum}. Examples are GW quantum states with no classical counterpart, analogous to nonclassical light in quantum optics. Ranging from squeezed \cite{guerreiro2020gravity, parikh2021b, cho2022quantum} to definite graviton number states \cite{tobar2024detecting}, nonclassical GWs could in principle leave signatures at detectors \cite{guerreiro2022quantum}, induce noise \cite{parikh2021b, cho2022quantum}, light-cone fluctuations \cite{ford1995gravitons, ford1996gravitons} and state-dependent gravitational entanglement and decoherence \cite{blencowe2013effective, de2015decoherence, oniga2016quantum, bassi2017gravitational, nandi2024quantum, xu2020gravitational, lagouvardos2021gravitational, kanno2021noise}. Measuring any of these effects would provide indirect evidence on the quantum nature of gravity, but this observational program is not met without challenges.
Unlike in electrodynamics, the quantization of matter does not imply quantization of gravity \cite{bronstein2012quantum, dyson2014graviton}. Moreover, in analogy to quantum optics \cite{clauser1974experimental}, the observation of gravitational wave noise or single graviton clicks in a detector are not sufficient to exclude a classical field-theoretic model of GWs \cite{carney2024graviton}. In order to determine the quantum nature of gravity via GW observations, we need to violate a nonclassicality witness. The possibility of doing this strongly relies on what quantum states of GWs can be produced in nature, which mechanisms generate them and how often they occur. Hence, we must turn to the question of \textit{sources} of quantum GWs. 

In quantum mechanics, nonlinearities lead to nonclassical states \cite{hillery1985conservation, coelho2009three, albarelli2016nonlinearity}. 
Gravity can be highly nonlinear \cite{scheel2014geometrodynamics}, so once again reasoning by analogy to quantum optics, we could expect that highly nonclassical GW states can be naturally prepared by the dynamics of strong gravitational fields. 
Strong nonlinearities could also act as efficient detectors for nonclassical GW states, since the output of a nonlinear interaction can be highly dependent on the quantum statistical properties of the input states, e.g. depending on whether they are bunched or anti-bunched \cite{shen1967quantum, agarwal1970field, kozierowski1977quantum, ekert1988second, olsen2002dynamical, qu1992photon, qu1995measurements, spasibko2017multiphoton}. To advance these ideas, we require a description of nonlinear, strongly interacting quantum GWs, which is in general a difficult problem as it involves quantizing gravity in situations where the spacetime curvature is large. We resort to perturbation theory.



We could imagine perturbing a curved dynamical spacetime with GWs and quantizing the perturbations by promoting their components and associated momenta to operators in Hilbert space and imposing canonical commutation relations. However, quantization requires the identification of observer-independent, physical degrees of freedom (DoFs) \cite{bergmann1956introduction, misner1957feynman, mandelstam1962quantization, donnelly2016diffeomorphism}, which are obscured by the right to choose arbitrary coordinate systems. There is in general no coordinate (gauge) independent way of characterizing perturbations of quantities in a spacetime, unless the Lie derivative of the unperturbed quantity with respect to any vector field vanishes identically in the background, which guarantees gauge invariance to linear order \cite{dewitt1965relativity, stewart1974perturbations, bruni1999observables}. Working with perturbed quantities that are exactly gauge invariant is therefore a too restrictive requirement, though there are situations where it is possible. 

One example where the quantization of GW perturbations can be successfully carried out in a curved dynamical background is the case of Robertson-Walker universes, where metric perturbations can be decomposed in scalar, vector and transverse-traceless (TT) tensor modes \cite{deser, straumann2008proof}. TT modes describe radiative degrees of freedom endowed with gauge invariance to linear order \cite{bardeen1980gauge, maggiore2018gravitational}. As shown by Ford and Parker, quantization of these radiative DoFs yields two scalar fields, one for each graviton polarization, minimally coupled to the spacetime background \cite{ford1977quantized}. This coupling leads to GW squeezed states in inflation \cite{grishchuk1993quantum, kanno2019nonclassical}, which are nonclassical according to the Glauber–Sudarshan P-representation \cite{hillery1985conservation}. Thus, rapid background expansion provides an example of a source of nonclassical GWs, but very likely inflation is an isolated event that occurred only once, if at all. 

Here, we aim at expanding the landscape of possible sources of quantum gravitational radiation by suggesting that nonclassical states can be efficiently produced when GW modes interact. To do this, we will relax the restrictive requirement of exact gauge invariance and instead work with approximate observer-independent quantities, by considering high-frequency GWs propagating in a weakly curved, i.e. low-frequency, GW background. Nonclassical features of both low- and high-frequency modes become prominent whenever the curvature associated to the waves is sufficiently strong, pointing towards the idea that appreciable quantum corrections might appear in strong field gravity. 

\textit{High-frequency waves.}--- Next, we review some points concerning high-frequency GWs. Readers familiar with \cite{isaacson1968gravitational1, isaacson1968gravitational} can skim this part and go directly to the argument for GW entanglement below. 

We will say that a GW perturbation has high frequency when its typical wavelength $\lambda $ is much shorter than the radius of curvature of the background spacetime $ \mathcal{R} $. As pointed out by Misner, in this limit, the projection operator onto the TT component of the wave becomes locally well defined to order $ \mathcal{O}(\lambda^{2}/\mathcal{R}^{2})$ \cite{misner64}.
This forms the basis for the high-frequency gravitational wave perturbation scheme developed by Isaacson \cite{isaacson1968gravitational1, isaacson1968gravitational}, who considered metrics of the form 
\begin{eqnarray}
    g_{\mu \nu} &=&  \gamma_{\mu \nu} + \epsilon h^{H}_{\mu \nu} \label{full_metric}
\end{eqnarray} 
where $ \gamma_{\mu \nu} $ describes the weakly curved background with radius of curvature of order $ \mathcal{R} $ and derivatives $ \partial \gamma_{\mu \nu} \sim \gamma_{\mu \nu} / \mathcal{R} $, $  h^{H}_{\mu \nu} $ is a metric perturbation containing components of typical wavelength $ \lambda $, amplitude $ \epsilon = \lambda / \mathcal{R} $ and $ \partial h^{H}_{\mu \nu} \sim h^{H}_{\mu \nu} / \lambda $. We assume $ \gamma_{\mu \nu} = \mathcal{O}(1)  ,\ h^{H}_{\mu \nu} = \mathcal{O}(1) $ and $  \epsilon \ll 1 $. Notice that while the amplitudes associated with the perturbations are small, their curvature can be quite high. 

A two-length scale expansion of the Riemann tensor in powers of $ \lambda $ and $ \mathcal{R} $ reveals the splitting
\begin{eqnarray}
    R_{\alpha\beta\gamma\delta}(g_{\mu \nu})  = R_{\alpha\beta\gamma\delta}^{(0)} + R_{\alpha\beta\gamma\delta}^{(1)} + R_{\alpha\beta\gamma\delta}^{(2)} + ...
    \label{riemann}
\end{eqnarray}
where $ R_{\alpha\beta\gamma\delta}^{(0)} \equiv R_{\alpha\beta\gamma\delta}(\gamma_{\mu\nu}) $ is the Riemann tensor of the background and the remaining terms quantify the contributions due to the metric perturbation in powers of $ \epsilon$. We refer to \cite{isaacson1968gravitational1} and the Supplemental Material for explicit expressions. Remarkably, the quantity $ R_{\alpha\beta\gamma\delta}^{(1)} $ has physical, observer-independent significance. Performing a coordinate transformation $ x^{\alpha} \rightarrow x^{\alpha '} = x^{\alpha} + \epsilon \xi^{\alpha} $ with $ \xi_{\alpha} = \mathcal{O}(1) $ and $ \xi_{\alpha;\beta} = \mathcal{O}(1) $ (semicolon denotes covariant differentiation with respect to $ \gamma_{\mu\nu} $), we find it transforms as $ R_{\alpha\beta\gamma\delta}^{(1)'} - R_{\alpha\beta\gamma\delta}^{(1)} = \pounds_{\xi}R^{(0)}_{\alpha\beta\gamma\delta} $, where $   \pounds_{\xi}R^{(0)}_{\alpha\beta} $ is the Lie derivative of the background curvature, with magnitude $ \mathcal{O}(\mathcal{R}^{-2}) $. Hence, instead of requiring that the Lie derivative of the unperturbed quantity vanishes exactly, the condition for exact invariance \cite{stewart1974perturbations}, we content ourselves with approximate invariance in the limit $ \epsilon \rightarrow 0 $. 



Denote the TT component of the perturbation as $ \bar{h}^{H}_{\mu \nu} $, which we recall is locally defined to $\mathcal{O}(\epsilon^{2})$ and satisfies $ \gamma^{\mu\nu} \bar{h}^{H}_{\mu\nu} = \bar{h}_{\mu\nu} ^{H \ ; \nu} = \bar{h}^{H}_{\mu 0} = 0 $. In the WKB, or geometric optics limit, we have $ h^{H}_{\mu\nu} = f e_{\mu\nu} e^{i\phi} $, where $ f $ is the wave amplitude, $ e_{\mu\nu} $ is the polarization tensor satisfying $ e_{\mu\nu}e^{\mu\nu} = 1 $ and $ k_{\mu} = \phi_{,\mu} $ is the wavevector. To leading order, the Einstein field eqs. imply $R_{\alpha\beta}^{(1)} = 0 $, which can be cast as as a wave equation (see Supplemental Material). Substituting the WKB expression in this wave eq. and separating the various terms according to their order in $ \epsilon $, we find $ k^{\nu}k_{\nu} = 0 $ to $ \mathcal{O}(\epsilon^{-2}) $ and
\begin{eqnarray}
     \ k_{\mu ; \nu}k^{\nu} = 0 \ , \ e_{\mu\nu ; \alpha} k^{\alpha} = 0 \ ,
    \label{transport_eqs}
\end{eqnarray}
to $ \mathcal{O}(\epsilon^{-1}) $, where once again corrections occur at $ \mathcal{O}(\mathcal{R}^{-2})$. Hence, high-frequency waves propagate through the background in the same way as electromagnetic radiation or any other kind of massless particles, with their polarization $ e_{\mu\nu} $ parallel transported along null geodesics generated by $ k_{\mu} $.

The second order term $ R_{\alpha\beta\gamma\delta}^{(2)} $ defines an effective stress-energy tensor $ T_{\mu\nu} $ for the high-frequency GWs via the next leading-order contribution to the Einstein eqs. $ R^{(0)}_{\alpha\beta} = -\epsilon^{2} R^{(2)}_{\alpha\beta} $.
Doing a Brill-Hartle (BH) average \cite{brill1964method} over many wavelengths of the wave and assuming the WKB approximation, $ T_{\mu\nu} $ reduces to the stress-energy tensor for null dust,
\begin{eqnarray}
    T_{\mu\nu}^{(\mathrm{BH})} \approx \mathcal{E} \hat{k}_{\mu}\hat{k}_{\nu} 
\end{eqnarray}
where $ \mathcal{E} = (\epsilon^{2}/64\pi G) k^{2} f^{2} $ is the energy density written in terms of the GW strain $ f $ and wavenumber $ k $, and $ \hat{k}_{\mu} = k_{\mu} / k  $.
If we have a pair of incoherent beams labeled $ 1 $ and $ 2 $ traveling along worldlines $ X_{j}(\lambda_{j}) $, their contribution to the total gravitational action is \cite{isaacson1968gravitational1}
\begin{eqnarray}
    S_{EH} \supset  \sum_{j = 1,2} \int d\lambda_{j} E_{j} \gamma_{\mu\nu} \frac{dX_{j}^{\mu}}{d\lambda_{j}}\frac{dX_{j}^{\nu}}{d\lambda_{j}}
    \label{particle_action}
\end{eqnarray}
where $ \hat{k}^{\mu}_{j} = dX^{\mu}_{j}/d\lambda_{j} $ and $ E_{j} = \int \mathcal{E}_{j} dV $ is a volume integral of the energy-density. Note that here, what we take to be null dust, or massless particles, arises from the geometric optics limit of high-frequency GWs propagating in the background spacetime, and appear as a consequence of the separation of scales within the Einstein-Hilbert action $ S_{EH} $. 


\textit{An argument for GW entanglement.}---
We are now in the position of arguing that within the effective field theory quantization of gravity, the nonlinearities inherent to the Einstein field eqs. can entangle GWs.

The essence of the argument relies on the fact that for a pure bipartite system $ AB $, the von Neumann entropy of the subsystems is a measure of the entanglement in the whole state, the entropy of entanglement \cite{schumacher1995quantum}. In particular, let $ AB $ start in a pure separable state and evolve according to some unitary dynamics. Then, if after some time $ A $ evolves to a mixture, $ AB $ must be entangled.

Consider the bipartite system $ AB $ to be comprised of high and low-frequency GW modes. Subsystem $ A $ consists of two beams of high-frequency waves, treated as bundles of massless particles in the geometric optics approximation traveling along neighboring worldlines $ X_{1}(\lambda_{1}) $ and $ X_{2}(\lambda_{2}) $. Assuming $ X_{1} $ and $ X_{2} $ are part of a congruence of geodesics, we take $ \lambda_{1} \equiv \lambda_{2}$. The DoF of $ A $ we are interested in is the geodesic deviation vector $ \xi = X_{2} - X_{1} $, which is a gauge invariant quantity and hence can be promoted to a quantum operator \cite{parikh2021b}. Subsystem $ B $ is composed by a low-frequency GW mode represented by a transverse-traceless tensor $ \bar{h}_{\mu\nu}^{L} $ of wavelength $ \mathcal{R}$. The background metric is
\begin{eqnarray}
    \gamma_{\mu\nu} = \eta_{\mu\nu} + \bar{h}_{\mu\nu}^{L}
\end{eqnarray}
and the low-frequency perturbation is canonically quantized as in \cite{parikh2021b}.

The time evolution of the total system is unitary and is given by the Einstein-Hilbert action, with the high- and low-frequency modes interacting through eq. \eqref{particle_action}. Writing this interaction in terms of the geodesic deviation vector $ \xi $ in Fermi normal coordinates \cite{bak2024quantumnull}, we find the coupling $ \int d\lambda \ \ddot{\bar{h}}_{ij}\xi^{i}\xi^{j} \subset	S_{EH} $. We may then consider the evolution of the initially separable state,
\begin{eqnarray}
    \vert \Psi(0)\rangle = \left( \vert \xi_{1} \rangle_{A} + \vert \xi_{2} \rangle_{A} \right) \vert \varphi \rangle_{B}
\end{eqnarray}
where $ \vert \xi_{i} \rangle_{A}$ are eigenstates of the geodesic deviation operator, and $ \vert \varphi \rangle_{B} $ is the initial state of the low-frequency GW perturbation. The reduced density matrix $ \rho_{A}(t) = \mathrm{Tr}_{B} \vert\Psi(t)\rangle\langle\Psi(t)\vert $ of the high-frequency GWs at time $ t $ can be computed by integrating out the low-frequency perturbation $ \bar{h}_{\mu\nu}^{L} $, following \cite{kanno2021noise, bak2024quantumnull}. In particular, the off-diagonal terms of $\rho_{A}$ evolve according to
\begin{eqnarray}
    \langle \xi_{i} \vert \rho_{A}(t) \vert \xi_{j} \rangle = \exp[-\Gamma(t)] \langle \xi_{i} \vert \rho_{A}(0)  \vert \xi_{j} \rangle \ , i \neq j 
    \label{off-diag}
\end{eqnarray}
where $ \Gamma(t) $ is the decoherence functional \cite{breuer2001destruction} (related to the Feynman-Vernon influence functional \cite{calzetta1994noise, anglin1997deconstructing}). In general, finding $ \Gamma(t) $ is not easy, but when $ \vert \varphi \rangle_{B} $ is a Gaussian state (such as a squeezed-coherent or a thermal state) it can be computed \cite{cho2022quantum}. The exact form of $ \Gamma(t) $ depends on $ \vert \varphi \rangle_{B}$, however, all we need for our argument is that it has a non-vanishing real part $ \Re(\Gamma(t)) > 1 $ for $ t > 0$ \cite{kanno2021noise, bak2024quantumnull, breuer2002}. 
As a consequence, the off-diagonal terms \eqref{off-diag} decrease, which is associated to a non-vanishing entropy of entanglement for subsystem $ A $ \cite{breuer2002}. Lo \& behold, high- and low-frequency GW degrees of freedom get entangled.

Therefore, we are led to conclude that if quantum particles undergo decoherence due to interactions with a quantized GW background (as shown in \cite{kanno2021noise, bak2024quantumnull}), then quantized GW modes must get entangled due to the nonlinearities present in the Einstein-Hilbert action. 

Within the linearized effective field theory quantization of GW perturbations, the potential of nonlinearities of Einstein gravity for generating non-classical effects is not manifest. The above argument settles this question, by showing that gravitational nonlinearities can, in fact, generate entanglement. We now turn to the question of whether non-classical effects can also appear within modes of similar frequency. 

\textit{Field interactions.}--- In addition to the which-path entanglement between high- and low-frequency modes, we can also expect that high-frequency waves propagating in a curved background can interact with each other to produce nonclassical states. The lagrangian density governing high-frequency GWs in a weakly curved spacetime can be written in terms of a pair of minimally coupled scalar fields,
\begin{align}
    L = \frac{1}{64\pi G}  \sqrt{-\gamma}   \gamma^{\mu\nu} \left( \bar{h}^{H}_{+,\mu} \bar{h}^{H}_{+,\nu} + \bar{h}^{H}_{\times,\mu} \bar{h}^{H}_{\times,\nu} 
 \right) 
 \label{total_density}
\end{align}
where $\bar{h}^{H}_{+} $ and $ \bar{h}^{H}_{\times}$ represent each of the graviton polarizations, and we made the substitution $ \epsilon \bar{h}^{H}_{+, \times} \rightarrow \bar{h}^{H}_{+, \times} $. The independence of \eqref{total_density} from the high-frequency polarization tensor can be physically understood as a consequence of the parallel transport eqs. \eqref{transport_eqs}, which tells us that the effects of the background curvature on the polarization occur at $\mathcal{O}(\mathcal{R}^{-2})$. This is analogous to the description of GW perturbations in Robertson-Walker spacetimes as two independent scalar fields \cite{ford1977quantized}.

From now on, we restrict ourselves to the $ + $ polarization of the high-frequency field for simplicity.
Expanding to linear order in $ \bar{h}_{\mu\nu}^{L} $ we have $ L = L_{0} + L_{I} $, where
\begin{eqnarray}
    L_{0} =  \frac{1}{64\pi G}  \eta^{\mu\nu} \bar{h}^{H}_{+,\mu} \bar{h}^{H}_{+,\nu} \label{free_lagrangian}
\end{eqnarray}
is the free lagrangian for high-frequency waves and
\begin{eqnarray}
    L_{I} = -  \frac{1}{64\pi G} \bar{h}^{L ij} \bar{h}^{H}_{+,i} \bar{h}^{H}_{+ ,j} \label{int_lagrangian}
\end{eqnarray}
describes a three-wave mixing GW process. In optics, three-wave interactions generally lead to nonclassicality and entanglement between the participating fields \cite{coelho2009three}. Can \eqref{int_lagrangian} produce nonclassical states?


To answer that, we decompose the low- and high-frequency fields in a discrete set of Fourier modes,
\begin{eqnarray}
    \bar{h}_{ij}^{L} &=& \frac{2\kappa}{\sqrt{V}} \sum_{\Vec{k},s} h^{L}_{s}(\Vec{k},t) e^{i\Vec{k}\cdot\Vec{x}} e^{s}_{ij}(\Vec{k}) \label{low_modes} \\
    \bar{h}^{H}_{+} &=& \frac{2\kappa}{\sqrt{V}} \sum_{\Vec{k}} h^{H}(\Vec{k},t) e^{i\Vec{k}\cdot\Vec{x}} \label{high_modes}
\end{eqnarray} 
where $ \kappa = \sqrt{8\pi G} $, $ e^{s}_{ij}(\Vec{k}) $ is the low field polarization tensor satisfying $ e^{s}_{ij}(\Vec{k}) e^{ij}_{s'}(\Vec{k}) = \delta^{s}_{s'} $ ($s = +, \times$) and we work in a cubic box of volume $ V = \ell^{3} $ with $ \Vec{k} = 2\pi \vec{n}/\ell $, $ \vec{n} \in \mathbb{N}^{3}$. Reality of the fields implies $ h^{L}_{s}(-\Vec{k})e^{s}_{ij}(-\Vec{k}) = h^{L}_{s}(\Vec{k})^{*}e^{s}_{ij}(\Vec{k})  $ and $ h^{H}(-\Vec{k}) = h^{H}(\Vec{k})^{*} $. 
Note that the sums in \eqref{low_modes} and \eqref{high_modes} range over different values. While for low-frequency modes $ \vert \Vec{k} \vert \lesssim
 2\pi/\mathcal{R} $, for high frequencies $  \vert \Vec{k} \vert \gtrsim 2\pi/\lambda $.
The low and high modes are quantized by imposing the standard commutation relations between the Fourier amplitudes $ h^{L}_{s}, h^{H} $ and associated momenta $ \Pi^{L}_{s} = \dot{h}^{L}_{s}  $, $ \Pi^{H} = \dot{h}^{H} $. It is convenient to define a pair of dimensionless hermitian high-frequency operators,
\begin{eqnarray}
    \mathbf{h}(\vec{k}) = \sqrt{\frac{k}{2}} \left( h^{H}(\vec{k}) + h^{H \dagger}(\vec{k}) \right) \ , \label{hermitian_field} \\
    \mathbf{\Pi}(\vec{k}) = \sqrt{\frac{1}{2k}} \left( \Pi^{H}(\vec{k}) + \Pi^{H \dagger}(\vec{k}) \right) \ , \label{hermitian_momentum}
\end{eqnarray}
which also satisfy canonical commutation relations  \cite{cohen1997photons}.


\begin{figure}[t]
    \includegraphics[scale=0.94]{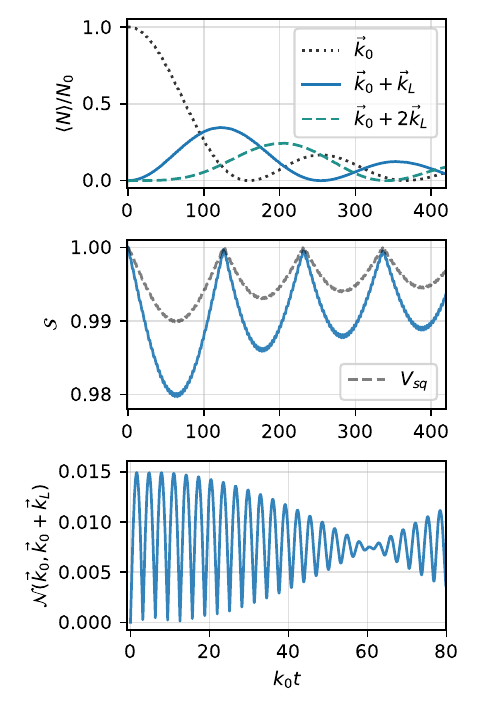}
    \caption{Nonlinear interactions as sources of nonclassical GWs. 
    Top: mean number of gravitons as a function of time for modes $ \vec{k}_{0},  \vec{k}_{0} + \vec{k}_{L} $ and  $\vec{k}_{0} +2\vec{k}_{L}  $. The initial number of quanta in mode $ \vec{k}_{0} $ is $ N_{0}= 10^{38}$. Middle: Squeezing degree $\mathcal{S}$ and squeezed variance $V_{\mathrm{sq}}$ of mode $ \vec{k}_{0} $. Bottom: Entanglement of modes $ \vec{k}_{0} $ and $ \vec{k}_{0}+\vec{k}_{L} $, quantified  by the logarithmic negativity $ \mathcal{N}(\vec{k}_{0}, \vec{k}_{0}+\vec{k}_{L})$. }
    \label{fig:squeezing_and_entanglement}
\end{figure}

The eqs. of motion for the Fourier components of $ \bar{h}^{H}_{+} $ are obtained by substituting the expansions \eqref{low_modes} and \eqref{high_modes} in the Lagrangian $ L $ and varying the action with respect to each $h^{H}(\vec{k})^{*}$. Together with the eqs. for the low-frequency modes $h^{L}_{s}(\vec{k})$, we obtain a system of nonlinearly coupled oscillators similar to the model proposed in \cite{yang2015coupled} for describing quasinormal modes of perturbed black hole spacetimes. Solving the complete nonlinearly coupled eqs. for low- and high-frequency modes is intractable. However, we may simplify the problem by assuming the \textit{undepleted pump approximation}, as customary in quantum optics. This consists of taking the background low-frequency GW to be in a coherent state and substituting the field operator $h_{s}^{L}(\Vec{k})$ by its expectation value. Considering as an example a high-frequency mode $ \vec{k_{0}} $ and a single-mode low-frequency background with wavevector $ \vec{k}_{L}$ orthogonal to $ \vec{k}_{0} $ ($ k_{L} \ll k_{0} $) and with polarization tensor aligned with the propagation direction of the high-frequency wave (see Supplemental Material for details), we have
\begin{align}
    \ddot{h}^{H}(\Vec{k}_{0},t) + k^{2}_{0} h^{H}(\Vec{k}_{0},t) = \nonumber \\
    -  q \left(  h^{H}(\vec{k}_{0} - \vec{k}_{L},t) + h^{H}(\vec{k}_{0} + \vec{k}_{L},t)  \right)
\end{align}
where $ q = \mathcal{A}k_{0}^{2} $ and we define the background amplitude $ \mathcal{A} = \alpha \sqrt{4\pi G / k_{L}V} $. The wave $ \vec{k}_{0} $ couples to additional high-frequency modes $ \vec{k_{0}} \pm \vec{k}_{L} $.

\begin{figure}[t]
    \includegraphics[scale=0.94]{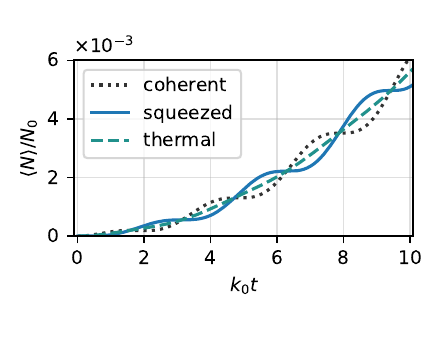}
    \caption{Nonlinear interactions as detectors of nonclassical GWs. Mean number of gravitons in mode $ \vec{k}_{0}+\vec{k}_{L}$ as a function of time for different initial states of mode $ \vec{k}_{0} $. All initial states have the same mean number of quanta $ N_{0} $.}
    \label{fig:detector}
\end{figure}

The quantity $ q $ can be thought of as a coupling constant governing the strength with which $\vec{k}_{0}$ interacts with $\vec{k}_{0} \pm \vec{k}_{L}$. Observe that the energy density of the background GW is $ \mathcal{E}_{L} = (1/32\pi G) k_{L}^{2} f_{L}^{2} $ where $ f_{L} $ is the background's strain, while the energy density of a single graviton of frequency $ k_{L} $ is $ \mathcal{E}_{g} = k_{L} / V $. The mean number of gravitons in the background is then $ N = \mathcal{E}_{L} / \mathcal{E}_{g} $, and the coherent state amplitude reads $\alpha = \sqrt{N} = f_{L} \sqrt{k_{L}V/32\pi G} $. We find the background amplitude is proportional to the low-frequency strain, $  \mathcal{A} = f_{L} / 2\sqrt{2} $. Since the order of magnitude of the Riemann curvature tensor associated to the high-frequency waves is $ \vert R_{\alpha\beta\gamma\delta}^{(1)}\vert  \approx \epsilon k_{0}^{2} $, we have
\begin{eqnarray}
 q \approx f_{L} \epsilon^{-1} \vert R_{\alpha\beta\gamma\delta}^{(1)}\vert  \ .
 \label{coupling_strength}
\end{eqnarray}
While $f_{L} $ is restricted to be much smaller than unity far from the source \cite{thorne1987gravitational}, it can achieve values on the order of a percent near cataclysmic events such as binary black hole mergers (see Supplemental Material for a discussion). Moreover, the magnitude of $ \vert R_{\alpha\beta\gamma\delta}^{(1)} \vert $ can be large \cite{isaacson1968gravitational1}, indicating that appreciable couplings can occur when the high-frequency curvature is strong. 


We can follow essentially the same steps discussed so far to derive the eqs. of motion for modes with wavevectors $ \vec{k}_{m} = \vec{k}_{0} + m\vec{k}_{L} $, where $m \in \mathbb{Z} $. In doing that, we find each of these couples to $ \vec{k}_{0} + (m+1)\vec{k}_{L} $ and $\vec{k}_{0} + (m-1)\vec{k}_{L}$ with the same strength $ q $, forming a system of linearly coupled quantum harmonic oscillators. Writing the eqs. of motion in terms of the dimensionless hermitian operators \eqref{hermitian_field} and \eqref{hermitian_momentum} we arrive at,
\begin{eqnarray}
    \dot{\mathbf{h}}(\vec{k}_{m}) &=& k_{m} \mathbf{\Pi}(\vec{k}_{m}) \nonumber \\
    \dot{\mathbf{\Pi}}(\vec{k}_{m}) &=& - k_{m} \mathbf{h}(\vec{k}_{m}) - \mathcal{A}k_{0} \left( \frac{\mathbf{h}(\vec{k}_{m + 1})}{\sqrt{\eta_{m}\eta_{m+1}}} + \frac{\mathbf{h}(\vec{k}_{m - 1})}{\sqrt{\eta_{m}\eta_{m-1}} }\right) \nonumber \\
    \ \label{bosonic_modes}
\end{eqnarray}
where $ \eta_{m} = 1 + m^{2} (k_{L}/k_{0})^{2} $ and $ k_{m} = k_{0} \sqrt{\eta_{m}} $. Note that as $ m $ increases, the \textit{coupling rate} $ \mathcal{A}k_{0}/\sqrt{\eta_{m}\eta_{m\pm1}} $ decreases, so higher frequencies effectively have less influence on the central mode $ \vec{k}_{0} $ and its neighbors $ \vec{k}_{0} \pm \vec{k}_{L} $, over a given period of time.

The dynamics of coupled bosonic oscillators is entangling \cite{plenio2004dynamics, brandao2021coherent}, and Gaussian states exhibit a non-vanishing entanglement for any value of the coupling rate \cite{audenaert2002entanglement}. This implies that \eqref{int_lagrangian} can indeed generate nonclassical states. To see that, take $ \vec{k}_{0} $ and $ \vec{k}_{0} + \vec{k}_{L} $ as the primary system, while the remaining modes will be regarded as part of the environment. We numerically simulate $ M $ coupled modes following eqs. \eqref{bosonic_modes} using tools from Gaussian quantum information \cite{weedbrook2012gaussian, brandao2022qugit}. Inspired by estimates of the peak GW strain emitted near binary black hole mergers, we consider $ \mathcal{A} = 0.015 $ and $ \epsilon = k_{L}/k_{0} = 10^{-4}$ (see Supplemental Material for a discussion). Mode $ \vec{k}_{0} $ is initially set in a coherent state  with a mean number of excitations $ N_{0} = 10^{38} $ \cite{dyson2014graviton}.
The remaining oscillators are initially in the vacuum state. For times $ t \leq 400 k_{0}^{-1} $, only the first few modes become appreciably populated, so we take $ M = 100 $ as an approximation. 

Fig. \ref{fig:squeezing_and_entanglement} shows the energy exchange and appearance of nonclassical features in high-frequency GWs. The top plot shows the exchange of the mean number of quanta between $ \vec{k}_{0}, \vec{k}_{0} + \vec{k}_{L} $ and $ \vec{k}_{0} + 2\vec{k}_{L} $ as a function of time. By means of interaction with $ \vec{k}_{0} $, the neighbor modes become appreciably populated within a cycle of oscillation. 
The transfer of energy between modes is related to the creation of massless particles in GW backgrounds \cite{grisaru1975gravitational, jones2017particle, su2017black, sawyer2020quantum}. In the middle plot, we show the \textit{squeezing degree} as a function of time, defined as the ratio of the squeezed $V_{\mathrm{sq}}$ to anti-squeezed $ V_{\mathrm{asq}} $ field quadrature variances $ \mathcal{S} = V_{\mathrm{sq}}/V_{\mathrm{asq}} $ \cite{brandao2022qugit}. Note that when $ V_{\mathrm{sq}} < 1 $ the state is nonclassical \cite{hillery1985conservation}. We see that a squeezed GW state with a macroscopic mean number of gravitons $ \gtrsim 0.2N_{0}$ is produced, with subsequent oscillations and gradual reduction of the squeezing degree, due to interaction with the environment, which consists of the remaining modes. The squeezing degree $ \mathcal{S} $ can be related to the squeezing parameter $ r $ according to $ \mathcal{S} = e^{-2r}$ \cite{davidovich1996sub}. Given the minimum value of $ \mathcal{S} \approx 0.98 $ attained in Fig. \ref{fig:squeezing_and_entanglement}, we find a maximum $ r \approx 10^{-2}$.

The bottom plot shows the evolution of the logarithmic negativity, a measure of entanglement \cite{weedbrook2012gaussian}. We see that $ \vec{k}_{0} $ and $ \vec{k}_{0}+\vec{k}_{L} $ display entanglement oscillations accompanied by a decay and subsequent revival caused by interaction with the remaining modes. This behavior is characteristic of linearly coupled quantum harmonic oscillators, and is predicted to occur in current tabletop levitated quantum optomechanics experiments; note also that the obtained negativity values are on the same order as for these optomechanical systems \cite{brandao2021coherent}. We have also performed numerical simulations considering smaller values of the amplitude $\mathcal{A}$ and perturbation parameter $\epsilon$ -- see Supplemental Material for details. In all simulations, a non-vanishing entanglement is observed in accordance with \cite{audenaert2002entanglement}.

Lastly, in Fig. \ref{fig:detector} we plot the mean number of quanta in mode $ \vec{k}_{0}+\vec{k}_{L} $ as a function of time for the cases in which the \textit{parent} mode $ \vec{k}_{0} $ starts in a coherent, squeezed or thermal state, all with the same initial mean number of gravitons $ N_{0} $. The rate at which higher order modes are populated depends on their parent's initial state. This is analogous to what occurs in nonlinear optics \cite{shen1967quantum, agarwal1970field, kozierowski1977quantum, ekert1988second, olsen2002dynamical, qu1992photon, qu1995measurements, spasibko2017multiphoton}, where the quantum statistics of the input radiation influences the efficiency of nonlinear conversion. Together with the idea that strong coupling can occur in highly curved situations (see eq. \eqref{coupling_strength}), this illustrates the potential of gravitational self-interactions as efficient detectors of quantum features of GWs.

\textit{Discussion.}--- We have shown that within the geometric optics limit, gravitational nonlinearities can in principle act both as sources and detectors of quantum GW states. The natural way forward is to transport these ideas to astrophysical situations relevant to GW observations. A possible setup, though likely not the only one, is the ringdown phase of black hole (BH) mergers. A perturbed BH during ringdown undergoes oscillations described by a family of exponentially damped sinusoids labeled by discrete angular $ (\ell,m) $ and overtone $ n $ numbers, with complex frequencies $ \omega_{(\ell,m,n)} $. The imaginary part of $ \omega_{(\ell,m,n)} $ sets the timescale for the modes' decay, implying that these quasinormal modes (QNMs) describe an intrinsically open system \cite{gardiner2004quantum}. Note that fast-spinning BHs tend to have a higher quality factor \cite{yang2013branching}.

Quantization of massless scalars, vectors and tensors, including QNMs in first order perturbation theory, has been discussed in the literature \cite{unruh1974second, candelas1981quantization, iuliano2023canonical}. An interesting case is that of a scalar field near a perturbed non-spinning BH, where coupling to QNMs produces squeezed states of the scalar \cite{su2017black}. At high orders (large $\ell$), geometric optics applies and we can think of QNMs as massless particles trapped in the light ring \cite{goebel1972, ferrari1984new, yang2012quasinormal} (see also \cite{fransen2023quasinormal, kehagias2024nonlinear}). This short wavelength limit of BH perturbations suggests a way of generalizing our argument for GW entanglement and the coupled oscillator model presented above, with low-order QNMs playing the role of a long-wavelength GW background and high-order modes approximately described by massless scalars. 

Beyond linear perturbations, nonlinear effects are ubiquitous in ringdown simulations \cite{zlochower2003mode, nakano2007second, london2014modeling, mitman2023nonlinearities, cheung2023nonlinear, redondo2024spin}. When a BH is perturbed by an incoming GW pulse with angular numbers $ (\ell,m) $ and amplitude $ A $, the output signal contains (i) ``additional modes with amplitudes scaling as powers of $ A $'', (ii) has ``significant phase shift and frequency modulation'', (iii) ``amplification'' and (iv) ``generation of radiation in polarization states not present in the linearized approximation'' \cite{zlochower2003mode}. Perturbed BHs act as a nonlinear element, where (i)-(iii) are characteristic of parametric amplification while (iv) hints that, contrary to the geometric optics case, polarization will matter in general. As observed by Mollow \& Glauber, the amplification of a field mode initially free of excitations can only occur by spontaneous emission processes, hence a theory of GW amplifications should be formulated in quantum mechanical terms \cite{mollow1967quantum}. The tools to describe nonlinear perturbations in a gauge invariant way exist \cite{bruni1997perturbations, campanelli1999second, gleiser2000gravitational, loutrel2021second}, hence such a quantum model of nonlinear self-interacting GWs around black holes can in principle be constructed. 

Equipped with a quantum theory of interacting modes, very general arguments can be drawn for the inevitability of nonclassical features \cite{caves1982quantum, hillery1985conservation}. Nonlinearities temper with particle number and field distributions, and classical theories cannot account for a great variety of those. 
Gravitational nonlinearities might turn out to be an important tool in experimentally settling the question on the quantum nature of spacetime. 
\\

\textbf{\textit{Note added:}} We have recently become aware that related ideas and a consistent squeezing parameter for quasinormal modes has been predicted in \cite{manikandan2025squeezed}.

\acknowledgments{
I acknowledge Antonio Zelaquett Khoury, George Svetlichny, Carlos Tomei, Igor Brand\~ao, Luca Abrahão, Felipe Sobrero and Maulik Parikh for conversations. I also express gratitude to Kip Thorne, for encouragement and for pointing out Isaacson's geometric optics approximation for gravitational waves. This work makes use of the QuGIT toolbox. We acknowledge support from the Coordenac\~ao de Aperfei\c{c}oamento de Pessoal de N\'ivel Superior - Brasil (CAPES) - Finance Code 001, Conselho Nacional de Desenvolvimento Cient\'ifico e Tecnol\'ogico (CNPq), Funda\c{c}\~ao de Amparo \`a Pesquisa do Estado do Rio de Janeiro (FAPERJ Scholarship No. E-26/200.251/2023 and E-26/210.249/2024), Funda\c{c}\~ao de Amparo \`a Pesquisa do Estado de São Paulo (FAPESP processo 2021/06736-5), the Serrapilheira Institute (grant
No. Serra – 2211-42299) and StoneLab.}

\appendix

\section{High frequency mode dynamics}

The Ricci tensor for high-frequency perturbations reads \cite{isaacson1968gravitational1}
\begin{eqnarray}
    R^{(1)}_{\mu\nu} = \frac{1}{2}\left(  h_{;\mu\nu} + h_{\mu\nu ; \sigma}^{\ \ \ \ \ \sigma} - h_{\sigma \mu ; \nu}^{\ \ \ \ \  \sigma} - h_{\sigma \nu ; \mu}^{\ \ \ \ \ \ \sigma} \right) \ , 
\end{eqnarray}
where covariant derivatives and index raising and lowering are done with respect to the background metric $ \gamma_{\mu\nu} $. To leading order ($\mathcal{O}(\epsilon^{-1})$) the Einstein field eqs. are $ \epsilon R_{\alpha\beta}^{(1)} = 0 $. Writing this in terms of the transverse-traceless high-frequency perturbation $ \bar{h}_{\mu\nu}^{H} $ we arrive at the wave equation,
\begin{eqnarray}
    \bar{h}^{H \ \ ;\beta}_{\mu\nu ; \beta} + 2 R^{(0)}_{\sigma\nu\mu\beta}\bar{h}^{H \beta\sigma} + R^{(0)}_{\sigma\mu} \bar{h}^{H \sigma}_{\nu} + R^{(0)}_{\sigma\nu}\bar{h}^{H\sigma}_{\mu} = 0 \ .
    \label{wave_eq}
\end{eqnarray}
Note this is the same expression appearing in (2.8) of \cite{ford1977quantized}. Writing $ \bar{h}_{\mu\nu}^{H} = \bar{h}^{H}_{s} \mathfrak{e}_{\mu\nu}^{s}$, where $ \mathfrak{e}_{\mu\nu}^{s} $ represent each of the high-frequency polarizations ($ s= + , \times$), we have \cite{isaacson1968gravitational1, isaacson1968gravitational}
\begin{eqnarray}
        \mathfrak{e}^{s}_{\mu\nu} \mathfrak{e}_{s'}^{\mu\nu} = \delta^{s}_{s'} \ , \ 
        k^{\mu} \mathfrak{e}^{s}_{\mu\nu} = 0  \ , \
        \\ \gamma^{\mu\nu}\mathfrak{e}^{s}_{\mu\nu} = 0 \ , \ 
        \mathfrak{e}^{s \ \ \ \  \sigma}_{\mu\nu ;\sigma} = \mathcal{O}(\mathcal{R}^{-2}) \ .
    \end{eqnarray}
Following essentially analogous derivations as in \cite{ford1977quantized}, neglecting second covariant derivatives of the polarization tensors, we arrive at the Lagrangian expression shown in eq. (9) in the main text, valid to $ \mathcal{O}(\mathcal{R}^{-2}) $. 

The eq. of motion for high-frequency modes is obtained by substituting the Fourier expansion for the high- and low-frequency fields in the Lagrangian $ L $ shown in the main text. Varying the total action with respect to $ h^{H}(\vec{k}_{0})^{*} $ yields the Heisenberg eq. of motion for a high-frequency mode with wavevector $ \vec{k}_{0} = k_{0}\hat{l} $, 
\begin{eqnarray}
    \ddot{h}^{H}(\Vec{k}_{0},t) + k^{2}_{0} h^{H}(\Vec{k}_{0},t) = \nonumber \\
    - \frac{\kappa}{\sqrt{V}} \sum_{\Vec{k}_{A}, \Vec{k}_{B},s} K_{s}(\Vec{k}_{A}, -\Vec{k}_{0}, \Vec{k_{B}}) h_{s}^{L}(\Vec{k}_{A},t) h^{H}(\Vec{k}_{B},t)
    \label{eq_high}
\end{eqnarray}
where,
\begin{eqnarray}
    K_{s}(\Vec{k}_{A},\Vec{k}_{0},\Vec{k}_{B}) = \delta(\Vec{k}_{A}+ \Vec{k}_{0} + \Vec{k}_{B}) e^{ij}_{s}(\Vec{k}_{A}) k_{0 i} k_{B j}
\end{eqnarray}
is analogous to the spectral function in nonlinear optics and $ e_{s}^{ij}(\vec{k}_{A})$ represent the polarization tensors of the low-frequency background GW perturbation.


As discussed in the main text, we make the \textit{undepleted pump approximation}, which consists of taking the background GW to be in a coherent state $ \vert \alpha \rangle $ and replacing the field operator $h_{s}^{L}(\Vec{k}_{A},t)$ by its expectation value. As an example, we take the wavevector of the background to be $ \vec{k}_{L} = k_{L} \hat{n} $ and its polarization $ e_{+}^{ij}(\hat{n}) = \hat{l}^{i} \hat{l}^{j} - \hat{m}^{i} \hat{m}^{j} $, with $ \hat{l}, \hat{m} $ orthogonal vectors satisfying $ \hat{l} \times \hat{m} = \hat{n} $. Observe that in doing this, we have fixed the orientation of the high-frequency wavevector with respect to the low-frequency polarization tensor. Note that this is simply an example and can be generalized to arbitrary orientations by considering different forms for $ e^{ij}_{s}$ and $ \vec{k}_{0}$. The low-frequency field expectation value reads,
\begin{eqnarray}
    \langle \alpha \vert  h^{L}_{s}(\Vec{k}_{A},t) \vert \alpha \rangle = \alpha u_{k_{L}}(t) \delta^{s}_{s'=+} \delta(\vec{k}_{A}-\vec{k}_{L}) \label{low_state1}
\end{eqnarray}
where $ u_{k_{L}}(t) = e^{ik_{L}t}/\sqrt{2k_{L}}  $ is the Minkowski mode function \cite{kanno2021noise} and $ \alpha $ is the coherent state amplitude, which for simplicity we take to be real. Via the field reality conditions, eq. \eqref{low_state1} implies the low-frequency field has both positive and negative wavevector components $ \pm \vec{k}_{L} $.
Substituting \eqref{low_state1} in \eqref{eq_high} and once again using the reality conditions, we find that the delta functions in $K_{s}$ enforce $ \vec{k}_{B} = \vec{k}_{0} \pm \vec{k}_{L} $, implying $ e_{+}^{ij}(\hat{n})k_{0 i} k_{B j } = k_{0}^{2}  $. Approximating the background as slowly varying in time, $ u_{k_{L}}(t) \approx 1/\sqrt{2k_{L}} $ ($ k_{L} \ll k_{0}, k_{B} $), we obtain the eq. of motion (16) in the main text.

\section{Strain estimates}


To motivate the amplitude value used in the main text, we now estimate the maximum peak strain $ h_{\rm peak} \sim \mathcal{A} $ near a binary black hole merger event. 

For asymptotically flat spacetimes, we define the \textit{wave zone} as a region where GWs propagate freely to infinity and independently from their source, i.e. perturbations have `wavelike' behavior \cite{arnowitt1961wave}. The wave zone occurs when $ k r \gg 1 $, or $ r \gg \lambda / 2\pi $, where $ k $ is the wavenumber of a GW with wavelength $ \lambda $ and $ r $ is the distance from the source. For a merger where both black holes have approximatelly the same mass, $ \lambda \sim R_{s} $, where $ R_{s} $ is the black holes' Schwarzschild radius. Henceforth, we take the wave zone at the end stages of a BH merger to begin at $ r \gtrsim 10R_{s} = 20M $, where in the last equality natural units are implied ($ G = c = 1$). Note that in numerical simulations of binary BH mergers, waveforms extracted at $r \gtrsim 25M $ are approximatelly free from gauge issues and consistent with wavelike propagation \cite{buonanno2007inspiral}.

We may then estimate the strain of the emitted GW using $ 
h \sim \frac{GM}{c^{2}r} \frac{v^{2}}{c^{2}} \  $\cite{maggiore2018gravitational},  
where $ v $ is the BH's relative velocity, $ M $ is the mass and $ r $ is the distance from the source. Plugging in $ r \sim 10R_{s} = 20M $, and $ v \sim 0.6 c$ -- the final relative velocity of the BHs in GW150914 \cite{abbott2016observation} -- we obtain $ h \sim 0.02$.

We can also estimate the peak strain by resorting to numerical simulations. The GW strain waveform $ h $ is commonly given in terms of $ r h / M $ (in natural units), which is approximatelly constant along outgoing null rays. The typical values of the peak strain for binary BH mergers similar to GW150914 is $ r  h_{\rm peak} / M \sim 0.1 - 0.2 $ (see \cite{mroue2013catalog} for the lower bound and \cite{yoo2023numerical} for the upper bound values). In the wave zone, this yields $  h_{\rm peak}  \sim 0.005 - 0.01$, consistent with the previous estimate.

We can also estimate the strain of GWs emitted from quasinormal modes near the source. Again from numerical simulations, we know the GW waveform far from the source at the end stages of a merger, during the ringdown phase, can be approximated by \cite{abbott2009search}
\begin{equation}
    h(t) \approx \mathrm{Re}\lbrace  A \sqrt{\delta E} \frac{GM}{c^{2}r} e^{i\omega_{22}t} \rbrace \ ,
\end{equation}
where $ \omega_{22} $ is the complex frequency of the dominant QNM with $ \ell = m = 2$, $ \delta E $ is the fraction of the black hole’s mass emitted in the form of GWs, $ r $ is the distance from the source, $ M $ is the final BH mass and $ A $ is a constant given by
\begin{eqnarray}
    A = \sqrt{\frac{5}{2}} Q^{-1/2} F(Q)^{-1/2} g(a)^{-1/2} \ ,
\end{eqnarray}
with $ a = J c / GM^{2}$ being the dimensionless angular momentum parameter corresponding to an angular momentum $ J $ (with $ 0 \leq a \leq 1 $) and
\begin{eqnarray}
    Q &=& 2 (1-a)^{-9/20} 
 \ , \\
    F(Q) &=& 1 + 7 / 24Q^{2} \ , \\
    g(a) &=& 1 - 0.63( 1- a )^{3/10} \ . 
\end{eqnarray}
The total fraction of the black hole's mass emitted in the form of GWs in the $ \ell = m = 2 $ mode lies between $ \delta E \sim 1\% - 3 \%$ \cite{goggin2008}. Putting the numbers together, for a non-spinning black hole we then find $ h_{\rm peak} = A\sqrt{\delta E} \frac{GM}{10c^{2}R_{s}} \sim 0.009 - 0.015 $, where the lower value corresponds to $ \delta E = 1\%$ and the upper value to $ \delta E = 3\%$. In view of these estimates, we choose the amplitude value $ \mathcal{A} = 0.015 $ used throughout the main text.

\begin{figure}[ht!]
    \includegraphics[scale=0.8]{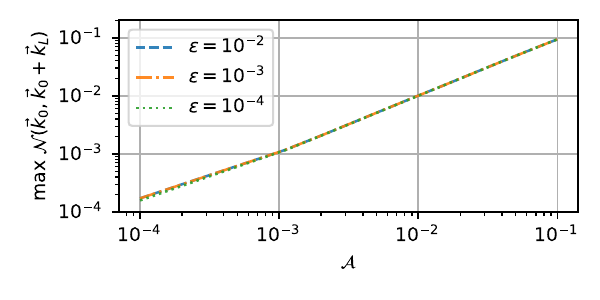}
    \caption{Maximum logarithmic negativity as a function of the amplitude $\mathcal{A} $ for different values of the perturbation parameter $ \epsilon$.  }
    \label{fig_params}
\end{figure}

We have also performed numerical simulations of GW entanglement generation for different values of the background strain amplitude $ \mathcal{A} $ and perturbation parameter $ \epsilon$. We have considered $ \mathcal{A} \in [10^{-4}, 10^{-1}]$ and $ \epsilon \in [10^{-5}, 10^{-2}]$, and computed the maximum logarithmic negativity for modes $ \Vec{k}_{0} $ and $ \Vec{k}_{0} + \Vec{k}_{L} $ as a function of the low frequency GW amplitude $ \mathcal{A} $, for different values of $ \epsilon$. The results are shown in Fig. \ref{fig_params}. We see that changing $ \epsilon $ within the chosen interval has very little effect on the generated entanglement. Moreover, entanglement is generated for any value of the coupling $ \mathcal{A} $. 

\bibliography{main}



\end{document}